\documentstyle[12pt]{article}
\newcommand{\sect}[1]{\section{#1}\setcounter{equation}{0}}

\input epsf
\begin{document}
\bigskip
\hskip 3.7in\vbox{\baselineskip12pt
\hbox{NSF-ITP-97-139}\hbox{hep-th/9711037}}
\bigskip\bigskip\bigskip\bigskip

\centerline{\large \bf Compactification in the Lightlike Limit}

\bigskip\bigskip
\bigskip\bigskip

\centerline{\bf Simeon Hellerman} 
\medskip
\centerline{Department of Physics}
\centerline{University of California}
\centerline{Santa Barbara, CA 93106}
\centerline{e-mail: sheller@twiki.physics.ucsb.edu}
\bigskip
\centerline{\bf Joseph Polchinski} 
\medskip
\centerline{Institute for Theoretical Physics}
\centerline{University of California}
\centerline{Santa Barbara, CA\ \ 93106-4030}
\centerline{e-mail: joep@itp.ucsb.edu}

\begin{abstract}
\baselineskip=16pt
We study field theories in the limit that a compactified dimension becomes
lightlike.  In almost all cases the amplitudes at each order of perturbation
theory diverge in the limit, due to strong interactions among the
longitudinal zero modes.  The lightlike limit generally exists
nonperturbatively, but is more complicated than might have been
assumed.  Some implications for the matrix theory conjecture are discussed. 
\end{abstract}
\newpage
\baselineskip=18pt

\sect{Introduction}

Matrix theory~\cite{bfss} is a promising proposal for the fundamental
degrees of freedom and Hamiltonian of M-theory.  The further
proposal~\cite{suss} of Susskind, which gives a physical interpretation to
the finite-$N$ matrix model, appears to be a major step forward.  This
proposal states that
\begin{equation}
\mbox{Finite $N$ matrix model}\ =\ 
\mbox{Discrete light cone quantization of M-theory} .
\label{fincon}
\end{equation}
The left-hand side of this equation has a precise definition in terms of
supersymmetric quantum mechanics, at least when the transverse dimensions
are noncompact.  It is the meaning of the right-hand side that we wish to
address.

Discrete light cone quantization (DLCQ) \cite{dlcq} refers to
compactification on a light-like circle,
\begin{equation}
(x^+, x^-, x^i) \cong (x^+, x^- + 2\pi R, x^i)\ , \label{nullbox}
\end{equation}
with fixed nonzero $p_- = n/R$.  For the purpose of the
conjecture~(\ref{fincon}), we believe that this must be understood as a
limit of compactification on spacelike circles.  This point of view has
also been taken in some very recent papers~\cite{bbpt}-\cite{deal}.

We should note that most of the literature on DLCQ is not directed toward
the above conjecture, but toward providing an infrared regulator for
light-cone quantized field theories. In this case the discrete theory has
no physical significance of its own, and the only physical criterion it
must satisfy is to give the correct infinite volume limit.  But for
conjecture~(\ref{fincon}) to be meaningful, the right-hand side must have
a natural and unique definition, and the limiting procedure provides
this.  If instead the DLCQ of M-theory is
something different, the conjecture loses much of its content and becomes
more of a definition.  Further, it implies that matrix theory has a whole
new moduli space of vacua, the discrete light-cone vacua, disconnected
from all previously known moduli spaces. We think that this is unlikely to
be true.

If our interpretation of the conjecture is correct, it raises a curious
point.  The finite-$N$ matrix theory is interpreted as one more limit
of M-theory.  But we already know many limits of M-theory: the various
string theories, and eleven-dimensional supergravity.  Why should one
more limit generate great excitement?  Presumably the answer is that
while matrix theory is a limit of M-theory, it is hoped that the full M-theory
can also be obtained as a limit of matrix theory, namely the limit of large
$N$: taking $N \to \infty$ at fixed $R$ is Lorentz-equivalent to
taking $R \to \infty$ holding the frame of an experiment fixed.

In this paper we will address not the second limiting procedure but
the first: does it make sense to put a quantum system in a
light-like box, in the limiting sense that we assume?  Our study in this
paper is limited to {\it quantum field theory,} rather than string or
M-theory. The purpose is in part to develop some intuition in this simpler
setting, but it is also of interest in its own right.  Lightlike
compactification is one of the few limits in which field theories
dramatically simplify, and therefore is a tool that should be developed
further.

We find that the situation is somewhat complicated.  
If we consider perturbation theory, the limit of lightlike
compactification does not exist.  That is, individual Feynman graphs
diverge due to the infamous zero modes.  In retrospect the problem is
rather obvious.  The zero modes are described by a field theory
in one fewer dimension, interacting with fixed degrees of freedom
representing the particles with nonzero $p_-$.  We are holding fixed the
parameters in the higher dimensional theory, so the coupling
$g^2$ of the reduced theory scales as $R_s^{-1}$ with $R_s$ being the
invariant length of the compact dimension.  One would therefore expect
every loop graph to diverge.  The only theories that have smooth limits
order-by-order are certain supersymmetric theories where the zero modes
interact with the fixed degrees of freedom but not with each other.

However, if we consider the full theory, then it is likely that the
limit does exist, at least if the original field theory itself exists
in the sense of being asymptotically free in the ultraviolet.  The
lightlike limit is governed by an {\it infrared} fixed point, and in
simple cases the zero modes simply become massive and cause no further
trouble.  While the limit appears to exist in most cases, our work points
up the fact that it is more complicated than expected.  

We should note that there are various discussions of zero modes in the DLCQ
literature, for example the recent paper~\cite{yama}.  However, because the
orthodox interpretation of DLCQ differs from ours, there seems to be little
relation between the treatments of the zero modes.  In particular, the
standard DLCQ appears to treat them essentially classically.\footnote
{For this reason we should perhaps introduce a new acronym, such as L$^3$
for the light-like limit, and restate the conjecture~(\ref{fincon}) as
`finite $N$ matrix model = L$^3$ of M-theory.'}

Although our work is not specifically applicable to M-theory, we include
in the conclusion some further discussion of recent work.

\sect{Scalar Field Theory}

We start with a complex scalar field theory in $d$ dimensions with
quartic self-interaction.  We will denote the time coordinate by $\tau$,
the periodic coordinate by $x^-$, and the remainder by $x^i$ for $i= 3,
\ldots, x^d$.  The metric and periodicity are
\begin{eqnarray}
ds^2 &=& -2 d\tau dx^- + \epsilon^2 dx^- dx^- + dx^i dx^i\ ,
\nonumber\\
(\tau,x^-,x^i) &\cong& (\tau,x^- + 2\pi R,x^i) \ .
\end{eqnarray}
The invariant length of the compact dimension is
\begin{equation}
R_s = 2\pi \epsilon R\ .
\end{equation}
The time $\tau$ is related to light-cone time $x^+$ by
\begin{equation}
\tau = x^+ + \frac{\epsilon^2}{2} x^- \ ,
\end{equation}
becoming identical in the limit $\epsilon \to 0$.

The action is
\begin{equation}
S = -\int d^dx \left(\partial_\mu \phi^* \partial^\mu \phi + M^2 \phi^*
\phi + \frac{g^2}{4} (\phi^* \phi)^2 \right)\ .
\end{equation}
We leave $d$ unspecified; a perturbatively
well-defined theory requires $d\leq 4$, a nonperturbatively
well-defined theory $d\leq 3$. 

Expanding
\begin{equation}
\phi(\tau,x^-,x^i) = (2\pi R)^{-1/2} \sum_{n=-\infty}^\infty
\phi_n(\tau,x^i) e^{i n x^-/R}\ ,
\end{equation}
the kinetic term is
\begin{equation}
\sum_{n= -\infty}^\infty \int d\tau\, d^{d-2}x^i 
\left( \epsilon^2 \partial_\tau \phi_n^* \partial_\tau \phi^{\vphantom x}_n
+ \frac{2in}{R} \phi_n^* \partial_\tau \phi^{\vphantom x}_n
- \partial_i \phi_n^* \partial_i \phi^{\vphantom x}_n 
-M^2 \phi^* \phi\right)
\ . \label{kinetic}
\end{equation}
This gives the propagator
\begin{equation}
\frac{i}{\epsilon^2 p_\tau^2 + 2n p_\tau/R - p_i p_i - M^2}\ =\
\frac{i}{\epsilon^2 p_\omega^2 - n^2 / \epsilon^{2} R^{2} - p_i
p_i -M^2} \ ,
\end{equation}
where $p_\omega \equiv p_\tau + n/\epsilon^{2} R$.

Now consider the one loop amplitude in figure~1,
\begin{figure}
\begin{center}
\leavevmode
\epsfbox{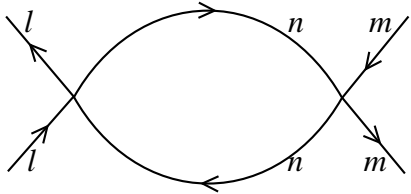}
\end{center}
\caption[]{One loop scalar graph.  Lines are labeled by $p_-R$, flowing
in the direction of the arrow.  The dangerous term is $n=0$.}
\end{figure}
\begin{eqnarray}
&&\frac{g^4}{4\pi^2 R^2} \sum_{n = -\infty}^\infty \int \frac{dq_\tau
d^{d-2}q_i}{(2\pi)^{d-1}} \label{loopamp}\\
&&\qquad
\Bigl( \epsilon^2 q_\omega^2 -  n^2 /\epsilon^2 R^2 - q_i q_i - M^2
\Bigr)^{-1}
 \Bigl( \epsilon^2 q'^2_\omega
- n^2 /\epsilon^2 R^2 - q'_i q'_i -M^2 \Bigr)^{-1} \ .\nonumber
\end{eqnarray}
Here $q'_\mu = q_\mu + k_\mu$, where $k_\mu$ is the exchanged momentum.
We have taken the dangerous case $k_- = 0$.  The problematic term is
$n = 0$, where both lines in the loop have vanishing longitudinal
momenta.  When $\epsilon = 0$, the integrand is independent of $q_\tau$ 
and the integral diverges.  The integral is proportional to $\epsilon^{-1}$
and diverges in the limit of lightlike compactification.

One way to understand this is in coordinate space.  In the lightlike
limit of $\epsilon = 0$, the kinetic term~(\ref{kinetic}) for the zero
modes has no $x^+$ (time) derivative and so the propagator is
proportional to $\delta(x^+)$.  A closed loop of zero modes then involves
$\delta(x^+)^2 \propto \delta(0)$.
We can also understand the divergence from dimensional reduction.  The
effective loop expansion parameter in the dimensionally reduced zero mode
theory is
\begin{equation}
g'^2 = \frac{g^2}{2\pi R\epsilon}\ .
\end{equation}
Leaving the external lines fixed and summing over all graphs with
internal zero mode lines reproduces the full complication of the
dimensionally reduced field theory, interacting with fixed sources
representing the external lines.  In the lightlike limit the coupling in
this theory diverges.

Now let us examine the case $d = 3$, where the original theory is weakly
coupled in the UV and should exist nonperturbatively.  
The effective dimensionless coupling at length
$l$ in the zero mode theory is $g^2 l^2 / R \epsilon$.  At the cutoff
distance $l = R \epsilon$ this actually goes to zero in the lightlike
limit, so the theory should be well-defined.  However, it diverges at any
fixed $l$.  In fact, one expects a mass gap at
\begin{equation}
l \approx \epsilon^{1/2} R^{1/2}/g
\end{equation}
where the effective coupling becomes strong.  Thus the zero mode
dynamics cures itself: at any fixed distance, the zero modes decouple
when $\epsilon$ is taken to zero.  However, amplitudes with vanishing $p_-$
exchange will be very different from their form in the noncompact theory,
due to the gap.

This same kind of analysis should apply to any theory that is
asymptotically free in the UV, such as four dimensional nonsupersymmetric
or supersymmetric gauge theory.  If the IR fixed
point has massless fields, then some residue of the zero mode dynamics
will survive.

Finally let us remark on the $n \neq 0$ terms in the
amplitude~(\ref{loopamp}).  At large $\epsilon$ these take the form
\begin{equation}
\frac{i g^4 \epsilon^2 R}{16 \pi^2 n^3} \int
\frac{d^{d-2}q_i}{(2\pi)^{d-2}} (1 + \epsilon^2 q_i q_i R^2 / n^2)^{-3/2}\ .
\end{equation}
At fixed $q_i$ this vanishes as $\epsilon \to 0$, consistent with the
analysis of Weinberg~\cite{weinberg}: for either time-ordering of the
vertices in figure~1, there is a particle with negative $p_-$. 
The integrated amplitude also vanishes in $d < 4$, but in $d = 4$ it scales
as $\epsilon^0$.  Further it is proportional to $n^{-1}$ so the sum
diverges.  This just reflects the fact that the nonzero modes renormalize
the coupling to the scale $R_s$.

\sect{The $+-0$ Model}

In the course of this investigation we did find one theory whose light-cone
limit is finite order-by-order.  This is a
toy model inspired by our eventual interest in $N=4$ gauge theories.  It is
a supersymmetric model with three chiral superfields and superpotential
\begin{equation}
W = g \Phi^+ \Phi^- \Phi^0\ .
\end{equation}
We also introduce nondynamical `Wilson lines,', so that the compact momenta
for $\Phi^\pm$ are shifted,
\begin{equation}
p_- = \frac{r}{R}\ ,\quad r \in {\bf Z} \pm \nu \ . \label{pshift}
\end{equation}
Here $\nu$ is an arbitrary noninteger constant.  The point is that the
fields in $\Phi^\pm$ do not have zero modes, so the effective zero mode
theory is free.

The absence of divergences is still nontrivial, because the zero modes
interact with the external states.
\begin{figure}
\begin{center}
\leavevmode
\epsfbox{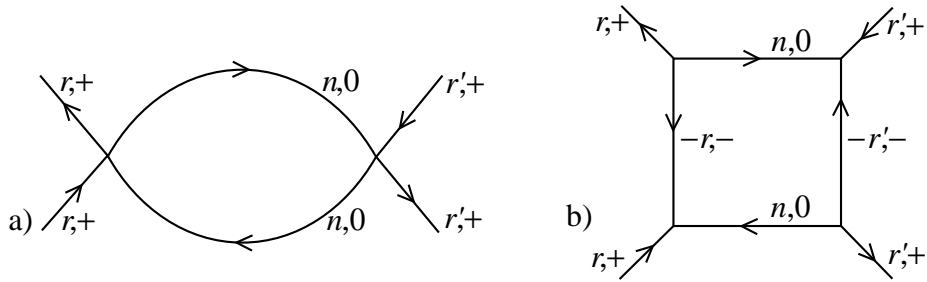}
\end{center}
\caption[]{a) A divergent scalar graph.  Lines are labeled by $p_-R$ and by the
superfield.  Other divergent graphs are obtained by replacing a pair of $+$
superfields with $-$. b) Canceling fermion loop.}
\end{figure}
Thus figure~2a has the same
divergence as figure~1.  Now, however, it is canceled by the fermion
loop of figure~2b.  More generally, any closed loop of bosonic zero modes,
generalizing figure~2a to $2M$ external lines, will give one net
$\delta(0)$.  The divergence is canceled by a corresponding fermion loop.

Rather than show the graphical calculations explicitly we give an argument
based on supersymmetry.  The zero mode theory lives at a single light-cone
time, so to first approximation we can ignore the $x^+$-dependence of the
external lines.  The supersymmetry transformations that close on
translation of $x^+$ are then unbroken by the external lines, and
guarantee net vanishing of the loop amplitude (again, to leading order in
$\epsilon$).  

Let us see this explicitly.  We write out in components the relevant terms
in the Lagrangian, using the conventions and notations of Wess and
Bagger~\cite{wess}, except that we denote scalars by $\phi$:
\begin{eqnarray}
{\cal L}&=&
-\,\epsilon^2 \partial_\tau \phi_0^{*} \partial_\tau \phi_0^{\vphantom x}
- \partial_i \phi_0^* \partial_i \phi^{\vphantom x}_0 
- i \bar \psi_0 \left(\bar\sigma^+ \partial_\tau +
\mbox{$\frac{1}{2}$} \epsilon^2 \bar\sigma^- \partial_\tau +
\bar\sigma^i \partial_i \right)
\psi_0 \nonumber\\
&&-\,i \bar \psi_r \left(\bar\sigma^+ \partial_\tau +
\mbox{$\frac{1}{2}$} \epsilon^2 \bar\sigma^- \partial_\tau +
i \bar\sigma^- r R^{-1} +
\bar\sigma^i \partial_i \right)
\psi_r \nonumber\\
&& -\, 2 \Lambda_r \psi_{-r} \psi_0 - {\rm h.c.}
- 4\Lambda^{*}_r \Lambda^{\vphantom x}_r \phi_0^* \phi^{\vphantom x}_0\ .
\end{eqnarray} 
We have kept only fields that contribute to the loop graphs.  In
particular, the scalars with nonzero $p_-$ do not appear, but the fermions with
nonzero $p_-$ appear in figure~2b.  All terms are quadratic in the quantum
fields.  We have adopted a rather condensed notation.  The subscript
labeling the superfields is omitted, because the $p_-$ moding is
sufficient to distinguish these.  The index $r$ runs over the values
(\ref{pshift}) for both $\pm \nu$, thus implying also a sum over the
superfields $\pm$.  The coupling and
backgrounds are joined in
\begin{equation}
\Lambda_r = g \phi_r\ .
\end{equation}

To make the $\epsilon  \to 0$ limit clear we redefine $\tau = \epsilon t$,
and make a Lorentz boost on the spinor indices so that
\begin{equation}
\sigma^+ \to \epsilon \sigma^+\ ,\quad \sigma^- \to \epsilon^{-1}
\sigma^-\ .
\end{equation}
The Lagrangian becomes
\begin{eqnarray}
{\cal L}&=&
-\, \partial_t \phi_0^{*} \partial_t \phi_0^{\vphantom x}
- \partial_i \phi_0^* \partial_i \phi^{\vphantom x}_0 
- i \bar \psi_0 \left(\bar\sigma^+ \partial_t +
\mbox{$\frac{1}{2}$} \bar\sigma^- \partial_t +
\bar\sigma^i \partial_i \right)
\psi_0 \nonumber\\
&&-\,i \bar \psi_r \left(\bar\sigma^+ \partial_t +
\mbox{$\frac{1}{2}$} \bar\sigma^- \partial_t +
i \epsilon^{-1} \bar\sigma^- r R^{-1} +
\bar\sigma^i \partial_i \right)
\psi_r \nonumber\\
&& -\, 2 \Lambda_r \psi_{-r} \psi_0 - {\rm h.c.}
- 4\Lambda^{*}_r \Lambda^{\vphantom x}_r \phi_0^* \phi^{\vphantom x}_0\ .
\end{eqnarray} 
The action acquires an overall $\epsilon$ from $d\tau$, which implies the
loop counting factor of $\epsilon^{-1}$.  Otherwise, $\epsilon$ appears
only in one term, where it causes one component of $\psi_r$ to decouple.
The surviving component, designated by a prime, satisfies
\begin{equation}
\bar\sigma^- \psi'_r = 0\ ,\quad \sigma^- \bar\sigma^+ \psi'_r
= 2 \psi'_r\ .\label{chiral}
\end{equation}

The Lagrangian finally comes to the form
\begin{eqnarray}
{\cal L}&=&
-\, \partial_t \phi_0^{*} \partial_t \phi_0^{\vphantom x}
- \partial_i \phi_0^* \partial_i \phi^{\vphantom x}_0 
- i \bar \psi_0 \left(\bar\sigma^+ \partial_t +
\mbox{$\frac{1}{2}$} \bar\sigma^- \partial_t +
\bar\sigma^i \partial_i \right)
\psi_0 \nonumber\\
&&-\,i \bar \psi'_r \bar\sigma^+ \partial_t 
\psi'_r  -\, 2 \Lambda^{\vphantom x}_r \psi'_{-r} \psi^{\vphantom x}_0 -
{\rm h.c.} - 4\Lambda^{*}_r \Lambda^{\vphantom x}_r \phi_0^*
\phi^{\vphantom x}_0\ . \label{finact}
\end{eqnarray} 
To leading order in $\epsilon$, the background is invariant under the
supersymmetry whose parameter $\xi'$ satisfies~(\ref{chiral}). 
Correspondingly the action~(\ref{finact}) is invariant under
\begin{eqnarray}
\delta \phi_0 &=& \sqrt 2 \xi' \psi_0 \ ,\nonumber\\
\delta \psi_0 &=& \sqrt 2i \sigma^i \bar\xi' \partial_i \phi_0
+ i\sqrt 2 \sigma^+ \bar\xi' \partial_t \phi_0\ ,
\nonumber\\
\delta \psi'_r &=& -2 \sqrt 2 \xi' \Lambda_{-r}^* \phi_0^{\vphantom x}\ .
\end{eqnarray} 
This acts linearly on the quantum fields, and so guarantees cancellation
of the leading $\epsilon^{-1}$ term in the one loop amplitude.

\sect{Discussion}

Our work points out that for rather simple dimensional reasons, the limit
of lightlike compactification leads to a strong coupling problem in almost
any field theory.  This may reduce the promise of this idea as a means of
studying field theory dynamics: DLCQ is not a free lunch.

To conclude, we comment on some very recent papers on matrix theory,
in particular one that seems to derive the matrix model~\cite{seilc}
and one~\cite{radine} that seems to show that it is incorrect. 
Other very recent papers~\cite{senlc,doog,deal} discuss related issues.

Roughly speaking, the recent paper by Seiberg~\cite{seilc} observes that
compactification on a nearly lightlike circle is Lorentz-equivalent to
compactification on a circle of small spacelike radius $R_s$.  The latter
compactification of M-theory gives the IIA string, but now in a sector
with nonzero D0-brane charge $p_- = n/R$.  Taking $R_s \to 0$ while
holding distances fixed in units of the eleven-dimensional Planck scale,
one retains just the open string ground states, which are indeed described
by the matrix theory.  This approach allows one to understand the
increase in the number of degrees of freedom when several coordinates are
periodically identified, from additional light states that survive the
limiting process.  Similar arguments are made by Sen~\cite{senlc}.

Dine and Rajaraman~\cite{radine} calculate a three-graviton to
three-graviton process in eleven-dimensional supergravity and obtain a
result that is not in agreement with the corresponding two-loop matrix
theory calculation.\footnote
{The paper~\cite{doog} of Douglas and Ooguri also reports a contradiction.
This in the context of a compactification of matrix theory, but is likely
closely related to the result~\cite{radine}.  The paper~\cite{ganor} of
Ganor, Gopakumar, and Ramgoolam also reports a contradiction, but this case
may be connected with the subtleties of compactification.}
Is this in direct contradiction with the
derivation in ref.~\cite{seilc}?  Does the argument in that paper actually
imply the previous successful tests of matrix model scattering, and therefore
that the matrix model and supergravity calculations in ref.~\cite{radine} must
agree?  We do
not see why this should be so.  The established range of validity for the
supergravity calculation is eleven large dimensions, while the
derivation of the matrix theory deals with M-theory compactified on a
circle small compared to the eleven-dimensional Planck scale.  Without
some additional physical input, mere boosts of coordinate systems and
uniform rescaling of units of length will not turn one regime into the
other.

We could try to provide the additional input as follows.  Consider the
supergravity scattering process in eleven large dimensions, but in a
frame (which can always be chosen for few enough particles) where
the $p_-$ components are integer multiples of some length $R$, assumed
to be greater than the eleven-dimensional Planck length.  One could then
consider the same process in a spacetime with the null
identification~(\ref{nullbox}).  Actually let us consider first an
identification that is almost null, with invariant periodicity $R_s$
much less than the Planck scale, and then take a limit.  By the
argument in ref.~\cite{seilc}, the resulting physics is indeed
described by the matrix theory Hamiltonian.  The one nontrivial step 
would then seem to be the periodic identification: should we expect this
to leave the amplitude invariant?

It is certainly not obvious that this should be so.
One effect of the compactification is that
loop momenta are quantized, leading as we have seen to strong coupling
effects.  The compactified theory then breaks down at longer distance, the
ten-dimensional rather than eleven-dimensional Planck scale.  A second
effect is the introduction of winding sectors, in this case
winding membranes which are IIA strings (this point is also made in
ref.~\cite{doog}).  This causes supergravity to break down at even longer
distance, the string scale. This is just the point that the supergravity
description is valid for small $R_s$ but distances large compared to the
string scale, while the matrix theory description is valid for small $R_s$ and
distances small compared to the string scale.

Thus, while the scaling argument of Seiberg shows that the 
conjecture~(\ref{fincon}) is literally true, it does not explain the
agreement with supergravity calculations, guarantee that future
supergravity calculations will agree or enable us to reconstruct the
eleven-dimensional limit.  For the same reason, the paper~\cite{bbpt}, which
purported to test the conjecture~(\ref{fincon}), does not.  Rather, it tests
some not yet clearly formulated assumption about continuation from the
supergravity regime to the matrix theory regime.

Additional input, perhaps the large-$N$ limit,
is needed.  Note that even at large $N$ the zero modes become strongly
coupled as $R_s \to 0$, so it is necessary to show that these decouple
from the large-$N$ process.

\subsection*{Acknowledgments} 

We would like to thank Melanie Becker, Katrin Becker, Shanta de Alwis,
Michael Dine, Eric Gimon, David Gross, Juan Maldacena, Hirosi Ooguri, Nati
Seiberg, and Charles Thorn for discussions and communications. 
This work was supported in part by NSF grants PHY94-07194
and PHY97-22022.

\end{document}